\documentclass[aps,prd,reprint,groupedaddress]{revtex4-2}
\usepackage{graphicx}
\usepackage{dcolumn}
\usepackage{bm}
\usepackage{hyperref}
\usepackage{amssymb}
\newcommand{\al}{\alpha}
\newcommand{\bt}{\beta}
\newcommand{\gm}{\gamma}
\newcommand{\dl}{\delta}
\newcommand{\ep}{\epsilon}

\newcommand{\et}{\eta}
\newcommand{\kp}{\kappa}

\newcommand{\ta}{\tau}

\newcommand{\ph}{\phi}

\newcommand{\ps}{\psi}

\newcommand{\Om}{\Omega}

\newcommand{\Lm}{\Lambda}

\newcommand{\Ph}{\Phi}

\newcommand{\half}{\frac{1}{2}}

\newcommand{\eela}[1]{\label{#1}\end{equation}}
\newcommand{\eeala}[1]{\label{#1}\end{eqnarray}}
\newcommand{\be}{\begin{equation}}
\newcommand{\ee}{\end{equation}}
\newcommand{\bea}{\begin{eqnarray}}
\newcommand{\eea}{\end{eqnarray}}

\newcommand{\Ft}{\tilde F}
\newcommand{\Gtata}{G^t_{\;\,t}}
\newcommand{\Gtar}{G^t_{\;\,r}}
\newcommand{\Grr}{G^r_{\;\,r}}
\newcommand{\Gthth}{G^\theta_{\;\,\theta}}
\newcommand{\Gphph}{G^\ph_{\;\,\ph}}
\newcommand{\Gmumu}{G^\mu_{\;\,\mu}}
\newcommand{\Gmunu}{G^\mu_{\;\,\nu}}
\newcommand{\gf}{y}
\newcommand{\fg}{f}
\newcommand{\fgi}{j}
\newcommand{\fref}{\gf_{\rm p}}
\newcommand{\rref}{r_{\rm p}}
\newcommand{\etmax}{\et_{\rm max}}
\newcommand{\amax}{a_{\rm max}}

\newcommand{\GN}{G_{\rm N}}

\newcommand{\mO}{\mathcal{O}}

\newcommand{\rmax}{r_{\rm max}}

\newcommand{\ellt}{\tilde\ell}

\begin{document}

\title{
Co\"{o}rdinate transformations, metrics and black hole features in the collapsed phase of EDT}
\author{Jan Smit\\
Institute for Theoretical Physics, University of Amsterdam, \\
Science Park 904, P.O.Box 94485, 1090 GL Amsterdam, the Netherlands.
}

\begin{abstract}
This is a companion article to {\em Using massless fields for observing black hole features in the collapsed phase of Euclidean dynamical triangulations} \cite{Smit:2023kln}.
It clarifies a singular co\"{o}rdinate transformation
of an $SO(4)$ invariant metric to the usual spherical co\"{o}rdinates in which, at an instant of time called zero, the metric takes the form of a black hole with an interior. Regular transformations are also studied and found to lead  in the zero time limit to the same spatial components of the metric as with the singular one, whereas the time component ends up differently.
Components of the Einstein tensor also end up the same.
A regular black hole metric is inversely transformed  and compared with simulation results in \cite{Smit:2023kln}.
\end{abstract}

\maketitle

\section{Introduction}
\label{secintro}

In Euclidean quantum field theory, configurations contributing to a lattice-regulated path integral are typically wildly varying on the lattice scale, whereas average propagators vary typically slowly. Likewise, in the Euclidean dynamical triangulation (EDT) approach to quantum gravity the simplicial configurations are wildly varying, but average scalar field propagators still vary slowly as a function of the geodesic lattice distance \cite{Agishtein:1991cv,deBakker:1996qf}.
In \cite{Smit:2023kln} we proposed using `measured' (numerically computed quantum-averaged) massless  propagators for defining an average metric. An intuitive idea supporting this is the fact that our experimental understanding of distance is essentially based on QED with its massless photon field.

In the collapsed phase of EDT, measurement of massless scalar field propagators led to the determination --- apart from an integration constant --- of the scale factor $a(\et)$  in an $SO(4)$ rotation invariant metric with line element
\be
ds^2 =
d\et^2 +  a(\et)^2 \, d\Om_3^2\,, \quad  0<\et<\etmax\,.
\label{4D}
\ee
Here $\et$ is a 
radial co{\"o}rdinate in four dimensions and $d\Om_3$ the line element on the $3$-sphere $S^3$ with unit radius; $a(\et)$ is defined to be positive. The scale factor was obtained in terms of a rational fit function
$f_{\rm rat}(\et)$:
\be
a(\et)= c_G\,  f_{\rm rat}(\et) = c_G\, \frac{p_0 +p_1 \, \et^2}{1+q_1\, \et^2}\,,
\label{frat}
\ee
where $c_G>0$ is the (dimensionless) integration constant.

\begin{figure}
\includegraphics[width=8cm]{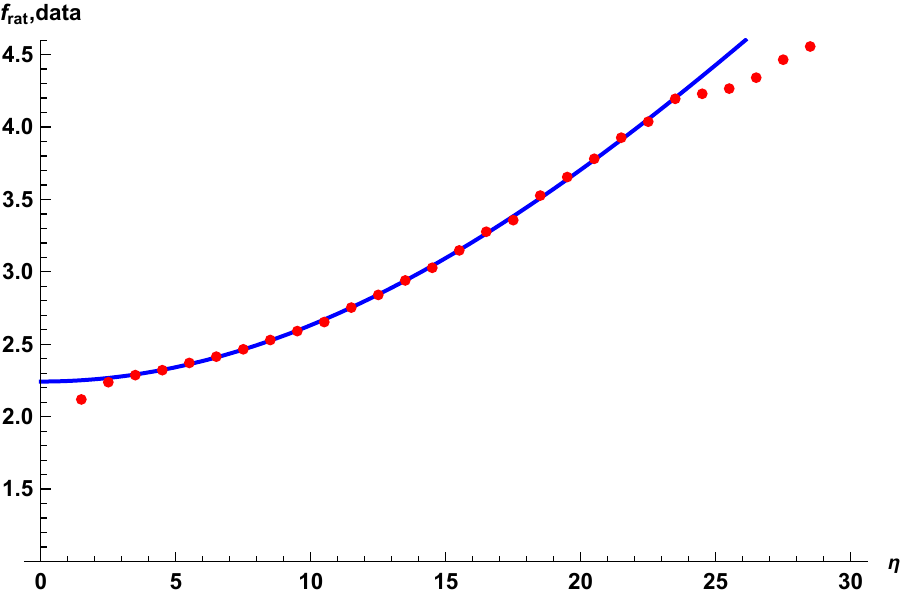} 
  \caption{
    Data (dots) and $f_{\rm rat}$ (blue curve); $p_0 = 2.2$, $p_1/p_0 = 0.0020$, $q_1=0.00022$. (data, $\et$ and $p_0$ in dual-lattice units $\ellt=1$;
$N_4=32$ k, $\kp_2=1.240$.)
  }
\label{figFitaG}
\end{figure}

Figure \ref{figFitaG} shows $f_{\rm rat}(\et)$ fitted to an example of measured data. The fit ignored data in $0<\et<5$ suspected to be too much influenced by lattice artefacts,  and also data in $\et> 24$ suspected of being susceptible to uncontrolled statistics, or finite size effects or `polymer hair'.

The scale factor is taken to represent a smooth continuum geometry with a three-ball boundary at the origin $\et=0$, of finite radius $a(0)$.
The constant $c_G$ gets determined when fitting to a continuum formula.

The slope $a'(\et) = da(\et)/d\et$  vanishes at the origin, $a'(0)=0$, which led to  black-hole features shown in \cite{Smit:2023kln}.
It may well be that more refined simulations on larger lattices suggest vanishing scale factors with non-vanishing slope at the origin, but which still have a (possibly nearly) vanishing slope somewhere away from the origin, which would then lead again to features of (possibly remnants of) black holes. This will be illustrated near the end of this article by the Hayward model \cite{Hayward:2005gi}.

After fitting numerical data for the metric in (\ref{4D}) a transformation to \emph{spherical co\"{o}rdinates} was presented in which the metrics are invariant under $SO(3)$ spatial rotations,
\be
ds^2 = g_{tt}(r,t)dt^2 + 2g_{rt}(r,t)dr\, dt + g_{rr}(r,t)dr^2 + r^2\, d\Om_2^2 .
\label{spherical}
\ee
Here $r$ is a radial co{\"o}rdinate in three dimensions, $t$ is a Euclidean time
variable and $d\Om_2$ is the line element on $S^2$ with unit radius. (In \cite{Smit:2023kln} imaginary time was denoted by $\ta$; here $t$ is used to avoid reading-confusion with $r$.)
The transformation was constructed for taking the limit $t\to 0$ where $g_{\mu\nu}$ became diagonal with $g_{tt}=1/g_{rr}$.

\begin{figure}
\includegraphics[width=7.9cm]{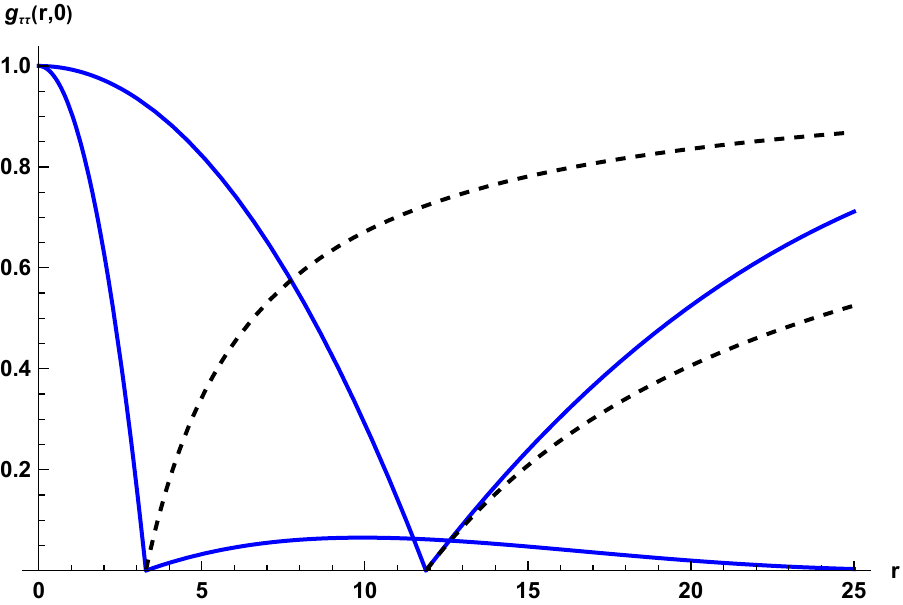}
\caption{Metric component $g_{tt}(r,0)=1/g_{rr}(r,0)$ for $c_G\simeq 1.5$ ($h\equiv c_G\, p_0\simeq 3.3$) and $c_G\simeq 5.3$ ($h \simeq 12$). The corresponding Schwarzschild metrics are also shown (black, dashed). (From \cite{Smit:2023kln}.)
}
\label{figgttS1240}
\end{figure}

Examples for several $c_G$ were given in \cite{Smit:2023kln}.
Figure \ref{figgttS1240}
shows $g_{tt}(r,0)$ for two cases of $c_G$. One recognizes features of a Euclidean black hole  \cite{Gibbons:1976ue,Hawking:1982dh} with horizon radius $h=a(0)=c_G\, p_0$ where $g_{tt}(r,0)=0$, and with an interior metric similar to `regular black holes' 
(\cite{Mazur:2004fk,Hayward:2005gi,Mazur:2015kia,Lan:2023cvz} and references therein).

The Einstein equations were assumed to hold in the effective sense and in the presence of a condensate of geometrical degrees of freedom.
The energy-momentum tensor of the condensate is determined by the Einstein tensor,
\be
G^\mu_{\;\,\nu} = 8 \pi \GN\, T^\mu_{\;\,\nu}\,,\quad G^\mu_{\;\,\nu} = R^\mu_{\;\,\nu} - \half\, R\, \dl^\mu_{\;\,\nu}\,.
\label{EinsteinT}
\ee
where $\GN$ is to be a Newton constant at the scale of $h$. In the limit $t\to 0$, $\Gmunu$ became also diagonal and examples were given for the various $c_G$.

In the present article we present calculations that led the construction of the co\"{o}rdinate transformation and the computation of the Einstein tensor in \cite{Smit:2023kln}. In section \ref{secinittrans} we introduce differential equations for a function $\gf(r,t)$ that serve to determine the co\"ordinate transformation and its effect on the form of the new metric.

In section \ref{secnumerical}, examples of $\gf(r,t)$ that produce diagonal metrics are calculated by numerically solving a differential equation for $\gf(r,t)$. When $t\to 0$ the inverse of the spatial component of the metric, $1/g_{rr}(r,t)$, is found to approach the curves shown in figure \ref{figgttS1240}. The time component $g_{tt}(r,t)$ behaves rather differently and its magnitude depends furthermore on time transformations $t\to\bar t(t)$ which are induced naturally by differences in the boundary condition for the differential equation. Its value at $t=0$ is ambiguous in this sense.
The Einstein tensor is also calculated using results in  \cite{Misner:1973prb} and it is found to become diagonal at time zero with unambiguous components $\Gmumu(r,0)$ (no summation) matching those in \cite{Smit:2023kln}.

In section \ref{secttozero} the work is done analytically. The diagonal-metric condition is given up but recovered in the limit $t\to 0$ together with $g_{tt}=1/g_{rr}$. It is shown how this leads to the function $\gf(r,t)$ and the ensuing metric in \cite{Smit:2023kln}; the Einstein tensor is recalled.

An inverse transformation is applied to the Hayward model in section \ref{secH}, where properties of the result are compared with the EDT data in figure \ref{figFitaG}.

Our conclusions are in section \ref{secconc}. Appendix \ref{appET} contains details of the Einstein tensor with a diagonal metric, appendix \ref{appEH} investigates an important cancellation in these formulas which prohibits a conjectured shell distribution at $r=h$ in \cite{Smit:2023kln}. The diagonality of $\Gmunu$ as a distribution is also investigated.

\section{Transformation to spherical co\"{or}dinates}
\label{secinittrans}

For convenience  metrics of the form (\ref{4D}) and (\ref{spherical}) are shown again more explicitly:
\bea
ds^2 &=&
d\et^2 + a(\et)^2\, d\ps^2 + a(\et)^2 \, \sin(\ps)^2 \, d\Om_2^2 \,,
\label{4D2}
\\
 &=& g_{tt}(r,t)\,dt^2 + 2g_{rt}(r,t)\,dr\, dt + g_{rr}(r,t)\,dr^2
\nonumber \\&& \mbox{}
+ r^2\, d\Om_2^2 \,,
\label{spherical2}
\\
d\Om_2^2 &=& d\theta^2 + \sin(\theta)^2\,d\ph^2\,.
\eea
The transformation is specified by a function $\gf(r,t)$ in
\be
r=a(\et)\,\sin(\ps)\,,\quad \gf(r,t)=a(\et)\,\cos(\ps)\,,
\ee
to be determined. In principle
\bea
0  &\leq&\et< \infty\,,\quad 0\leq\ps\leq\pi\,,
\\
 -\infty &<&t <\infty \,,\quad 0\leq r < \infty\,,
\eea
in practice $0\leq\et\leq\etmax$ with corresponding limits on $r$ and $t$ depending on $\gf(r,t)$.
Since $a>0$,
\bea
y &>& 0\,, \quad 0<\ps<\pi/2\,,
\\
y &<& 0\,, \quad \pi/2<\ps<\pi\,.
\eea
Note
\be
r^2 + \gf^2 = a^2\geq h^2\,,\quad h\equiv a(0)\,,
\label{rsqplusfsq}
\ee
which will be used in the following; the case with $a=h$ will be called \emph{the boundary equation}.
In the region where $a(\et)$ is a monotonously increasing function of $\et$ it has a unique inverse $\et(a)$, and a positive function $F(a)$ can be defined by
\be
(d a/d \et)^2=F(a) = F\left(\sqrt{r^2 + \gf^2}\right).
\label{introF}
\ee
The new metric turns out as
\bea
g_{tt} &=& \frac{\dot \gf^2}{r^2 + \gf^2}\left(\frac{\gf^2}{F} + r^2\right)\,,
\label{gtata}
\\
g_{rt} &=& \frac{\dot \gf}{r^2 + \gf^2}\left(\frac{\gf(\gf \gf^\prime+r)}{F}-r(\gf-r \gf')\right)\,,
\label{grta}\\
g_{rr} &=&\frac{1}{r^2 + \gf^2}\left(\frac{(\gf \gf'+r)^2}{F}+(\gf-r \gf')^2\right)\,
\label{grr}
\eea
($\dot \gf = \partial_t \gf$, $\gf'=\partial_r \gf$).
We would like $\gf$ to be such, that the off-diagonal component of the metric vanishes,
\be
g_{rt}= 0\,.
\label{eqncross}
\ee
Equation (\ref{eqncross}) can be solved for $F$ to eliminate it from expression (\ref{grr}) for $g_{rr}$ and obtain
\be
g_{rr} = 1-r\,\gf^\prime/\gf\,,
\label{grrffp}
\ee
which is useful once $\gf$ that satisfies (\ref{eqncross}) is known.
Assuming that generically $\dot\gf \neq 0$, solving (\ref{eqncross}) for $\gf^\prime$ gives
\be
\gf^\prime= r \gf \frac{F-1}{\gf^2 + r^2 F}\equiv P\,,
\quad P=P(r,\gf)\,.
\label{fprime}
\ee
We also would like $\gf$ to satisfy the property
\be
g_{tt} = 1/g_{rr} \,,
\label{eqnggt}
\ee
as for the Schwarzschild Euclidean (Anti) de Sitter
(SE(A)dS) metrics \cite{Hawking:1982dh}. Requiring (\ref{eqnggt}) in addition to 
(\ref{eqncross}) leads to an equation for
$\dot\gf^2$, or
\be
\dot\gf = \pm\sqrt{F}\,.
\label{fdot}
\ee
The case with a minus-sign is the time-reversed version of the case with a plus-sign.
When using this equation in the following we choose the plus-sign to avoid double covering.

The differential $d\gf = P\, dr + \sqrt{F}\, dt$ is in general {\em imperfect},
i.e.\ loosely $\partial_t P \neq \partial_r \sqrt{F}$, meaning 
\bea
\partial_\gf P(r,\gf)\,\sqrt{F(\sqrt{r^2 + \gf^2})}
&\neq& \partial_r \sqrt{F(\sqrt{r^2 + \gf^2})}
\\
&& \left. + \partial_\gf \sqrt{F(\sqrt{r^2 + \gf^2})}\, P(r,\gf) \right. .
\nonumber
\eea
Requiring the differential to be perfect leads to a differential equation for $F$ which is easy to solve,
\be
dF/da=2(F-1)/a\; \Rightarrow\; F=1+ c_F\, a^2 = 1 + c_F (r^2 + \gf^2)\,,
\ee
where $c_F$ is an integration constant. Writing $c_F=\pm 1/r_0^2$ with $r_0>0$,
the solution of (\ref{fdot}$^+$) with initial condition $\gf(r,0)=0$ is,
\bea
\gf(r,t) &=&\sqrt{r_0^2 + r^2}\, \sinh\frac{t}{r_0}\,, \qquad\mbox{(EAdS)}
\label{fEADS}
\\
\gf(r,t) &=&\sqrt{r_0^2-r^2}\, \sin\frac{t}{r_0}\,, \quad\qquad\mbox{(EdS)}
\label{fEDS}
\eea
respectively for $+$ and $-$. In the latter case $r$ is to be limited to $0<r<r_0$.
Upon substitution in (\ref{gtata}) --
(\ref{grr}) the time-dependence drops out and the diagonal EAdS and EdS metrics emerge,
\be
g_{tt}(r) = 1 \pm r^2/r_0^2\,.
\label{E(A)dS}
\ee
Furthermore, treating (\ref{introF}) as a differential equation for the scale factor $a(\et)$ leads for the $+$ case to the solution $a(\et)=r_0 \sinh[(\et-s_0)/r_0]$ with integration constant $s_0$ (or its $\et$-reversed version), corresponding to hyperbolic space.
In the $-$ case it leads to the spherical scale factor  $a(\et)=r_0 \sin[(\et-s_0)/r_0]$.

Integrating the imperfect differential $d\gf$ along a path in the $(r,\,t)$ plane gives a path-dependent result. To avoid this imperfection we shall in section \ref{secnumerical} release the condition $g_{tt}=1/g_{rr}$.

\subsection{Regular transformation to diagonal metric}
\label{secnumerical}

In this section the condition of diagonality, $g_{rt}=0$ is kept and the differential equation (\ref{fprime}) will be integrated at fixed times. 
Equation (\ref{fdot}) will be used only at one reference $r$ to obtain boundary conditions depending on time for the integration of (\ref{fprime}) along $r$. Then there is no imperfectness of the  co{\"o}rdinate transformation.

We concentrate on the fit function (\ref{frat}). It is convenient to rewrite its $a(\et)$ in the form
\be
a(\et) = h\,\frac{1+p\,\et^2}{1+q\,\et^2}\,,\quad
h=c_G\, p_0\,,\; p=\frac{p_1}{p_0}\,,\; q=q_1\,.
\label{apq}
\ee
The function $F(a)$ introduced in (\ref{introF}) can be determined via the inverse function $\et(a)$ of $a(\et)$; one finds
\bea
\et^2&=& \frac{a-h}{p h - q a},\,\quad
a^\prime(\et)^2 = \frac{4 h^2 (p-q)^2 \et^2}{(1+q\,\et^2)^4}\,
\label{eta}
\eea
and
\be
F(a)= \frac{4 q^3(a-h)(h p/q - a)^3}{h^2 (p-q)^2}\,.
\label{Frat}
\ee
The function $F(a)$ is positive between its zeros at $a=h$ and $a=h\,p/q$ with a maximum at $\amax$ somewhere in between, and only its monotonous branch in $h<a<\amax$ is to be used. In the numerical simulation \cite{Smit:2023kln} $\amax$ was only slightly larger than $2h$. In this section results will be shown for the generic case $c_G=1.5$ ($h=3.3$) (cf.\ figure \ref{figgttS1240}).

Since an analytical treatment is already awkward in this case and might be prohibitive with more general fit functions, the following is a numerical exploration.

We choose a reference distance $\rref$ and erect a `time pole' $\fref(t;\rref)$ by solving $\dot \gf=\sqrt{F}$ numerically at $r=\rref$:
\be
\dot \fref(t;\rref)=\sqrt{F\left(\sqrt{\rref^2 + \fref(t;\rref)^2}\right)}\,,
\label{eqnpole}
\ee
with initial conditions $\gf=\gf_0$ at $t=t_0$ chosen as follows.
Guided by (\ref{fEADS}) we assume that $\gf>0$ ($\gf<0$) when $t>0$ ($t<0$). Compatible with this, the initial condition can be taken a minimal $|\gf|$ at $t=0$: for $\rref >h$, $\gf_0=0$, for $\rref<h$  the minimal $|\gf|$ has to comply with the boundary equation in (\ref{rsqplusfsq}):
\bea
\gf_0 &=& 0\,,\qquad\quad\quad  h \leq \rref \leq\rmax\,, \quad t_0 =0\,,
\nonumber\\
&=& \pm \sqrt{h^2-\rref^2}\,,\quad 0\leq \rref \leq h\,,\quad t_0 \to 0^\pm\,.
\label{initialcond}
\eea
The resulting $\fref$ is a monotonically rising function of $t$.
Next, equation  (\ref{fprime}) implementing $g_{rt}=0$ is solved (numerically) with a boundary condition attaching $\gf$ to the pole:
\be
\gf^\prime(r,t;\rref)=P(r,\gf(r,t;\rref))\,,\quad \gf(\rref,t;\rref) = \fref(t;\rref)\,.
\label{attpole}
\ee
The solutions $\gf(r,t;\rref)$ map the $(r,t)$ plane to the $(r,\gf)$ plane and the give a foliation of the latter. The foliation lines do not cross.

Consider erecting a second time pole
$\bar\fref(\bar t;\bar\rref)$
at a different position $\bar\rref\neq \rref$. This pole will be crossed by a foliation line $\gf(r,t;\rref)$ of the first pole,
$\gf(\bar\rref,t;\rref)=\bar\fref(\bar t;\bar\rref)$, which determines $\bar t$ in terms of $t$. This can be interpreted as a co{\"o}rdinate transformation of the time variable only:
$\bar t=\bar t(t)$.
The second pole leads to a second foliation
$\gf(r,\bar t;\bar\rref)$. However, substituting $\bar t=\bar t(t)$ will give just the original foliation.

In other words: results of different choices of the pole position are related by  co{\"o}rdinate time transformations that depend only on time. We shall see evidence of this in tensor components that transform as a scalar field under such transformations, in particular
the spatial components of the metric (which is diagonal in this section)
and the diagonal up-down components of the Einstein tensor, $G^\mu_{\;\,\mu}(r,t)$ 
(no summation).

Results will be shown for two choices of $\rref$ \& initial conditions for $\fref(t;\rref)$:
\bea
\rref &=& 2 h \; \&\; t_0=0\,,\;\fref(0;2 h)= 0\,,
\label{condpole}
\\
\rref &=&0\; \&  \;\fref(t_0;0)= \pm h\,(1 + c_t t_0^2)\;,\;
c_t=p-q\,,
\label{condpolein}
\eea
with very small but nonzero $c_t t_0^2>0$ to get the numerical integration started. (With the indicated choice of $c_t$, the solution of (\ref{eqnpole}) at small times is $\pm h(1+ c_t\, t^2 + \mO(t^4))$, cf.\ section \ref{secttozero}.)
The case of $-h$ in (\ref{condpolein}) corresponds to negative $t$, $\fref(t;\rref)$ and $\gf(r,t;\rref)$.

\begin{figure}
\includegraphics[width=8cm]{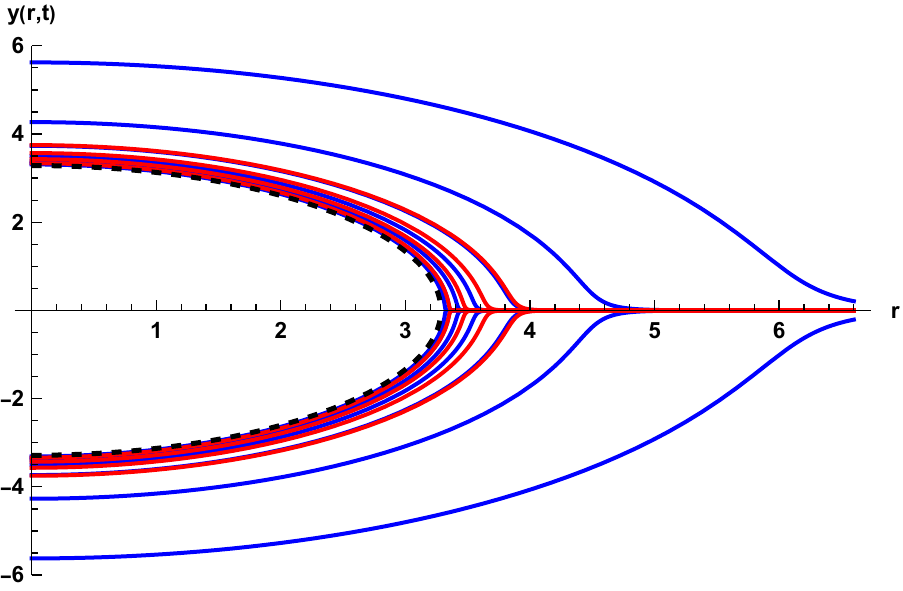} 
\caption{
Foliation curves $\gf(r,t)$ for $\rref= 2h$ and $t= \pm 10^{k-24}$, $k=0,4,8,\cdots,24$ (blue), and for $\rref=0$ at times $t= \pm \{3,5,7,9\}$ (red, partially overlaying the blues curves). The inner curve (dashed, black) represents $\pm\sqrt{h^2 - r^2}$; $h=3.3$ .
}
\label{figfol}
\end{figure}

\begin{figure}
\includegraphics[width=8cm]{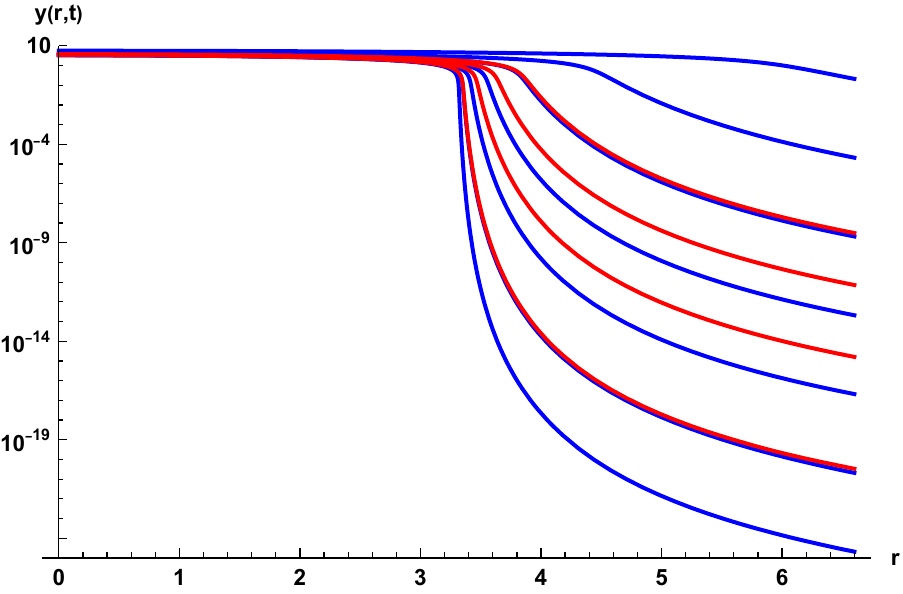} 
\caption{
Logarithmic plot corresponding to the positive $\gf$ part of figure \ref{figfol}.
}
\label{figfollog}
\end{figure}

Figure \ref{figfol} shows foliation curves obtained with the two choices of $\rref$ in (\ref{condpole}), (\ref{condpolein}). The curves run over the whole domain $0<r<\rmax$ and approach the same limit (dashed, black) when $t\to 0$.
Figure \ref{figfollog} shows a corresponding logarithmic plot.

The calculation of the metric needs derivatives of $\gf(r,t)$. Spatial derivatives can be expressed as functions of $\gf$ without derivatives using the first equation in (\ref{attpole}), repeatedly as needed, for example
\be
\gf^{\prime\prime} = \partial_r P(r,\gf) + \partial_y P(r,\gf)\, P(r,\gf)\,.
\ee
Time derivatives of $\gf$ were calculated using nearby foliations at $t\pm \ep$ and
\bea
\dot\gf(r,t)&\simeq& [\gf(r,t+\ep)-\gf(r,t-\ep)]/(2\ep)\,,
\label{dotsy}
\\
\ddot \gf(r,t)&\simeq& [\gf(r,t+\ep)+\gf(r,t-\ep)-2\,\gf(r,t)]/\ep^2\,,
\nonumber
\eea
with small $\ep$ typically of order $t/10$.
(Evaluating $\gf$ also at $t\pm 2 \ep$ one can approximate $\dddot\gf$ and also improve the above approximations. The first derivative $\dot\gf$ is needed for $g_{tt}$, but it turns out that the time derivatives cancel out of components of the Einstein tensor.)

\begin{figure}
\includegraphics[width=8cm]{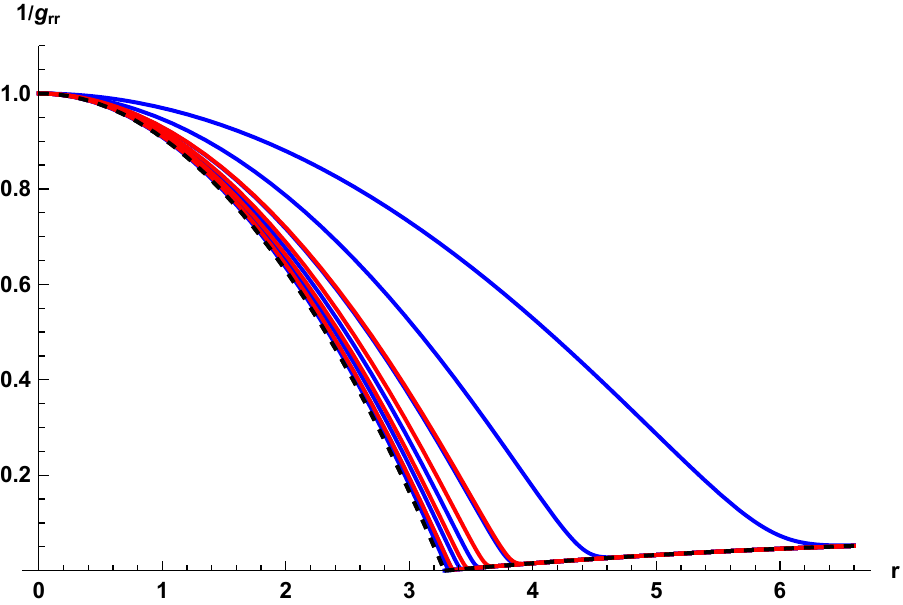} 
\caption{
Plot of $1/g_{rr}$. The dashed (black) curves represent $F(r)$ in the exterior and $1-r^2/h^2$ in the interior.
}
\label{figginvall}
\end{figure}

\begin{figure}
\includegraphics[width=8cm]{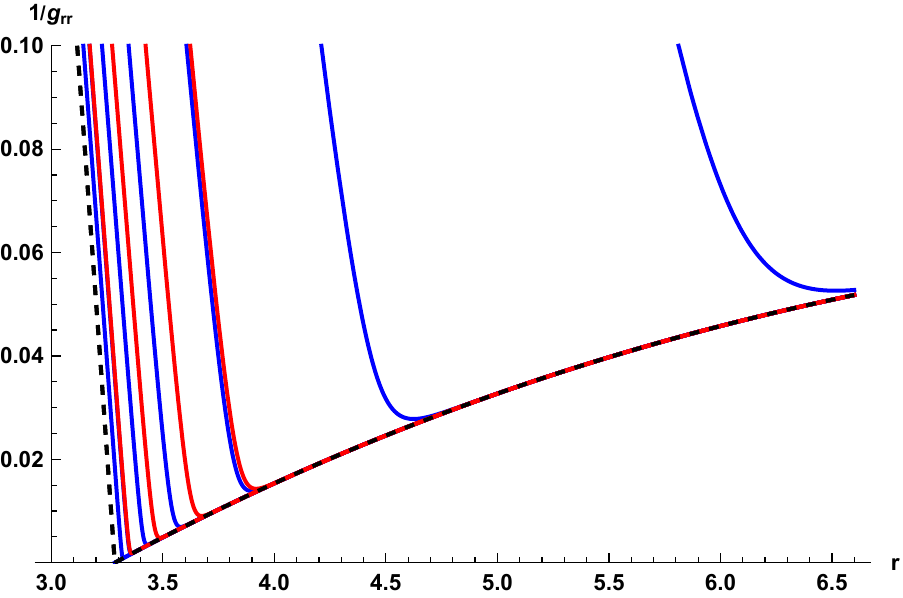} 
\caption{
Closeup of figure \ref{figginvall}. The envelope covers visually the $1/g_{rr}$ curves the exterior region; it is zero at $r=h$.
}
\label{figginv}
\end{figure}

Figure \ref{figginvall} shows $1/g_{rr}(r,t)$ obtained with the help of (\ref{grrffp}). As $t$ approaches zero an envelope develops, which is represented by the dashed black curves. Figure \ref{figginv} shows a closeup.
The envelope is $F(r)$ in the exterior region $r>h$ and switches to $1-r^2/h^2$ in the interior. In the exterior this can be understood from taking the limit $\gf\to 0 $ and $\gf' \to 0$ in the basic form (\ref{grr}) for $g_{rr}$. In the interior the limit $\gf\to \pm \sqrt{h^2-r^2}$ leads to $F(h)$ in the denominator in (\ref{grr}) and one has to heed the fact that $F(h)=0$ (cf.\ section \ref{secttozero}).

\begin{figure}
\includegraphics[width=8cm]{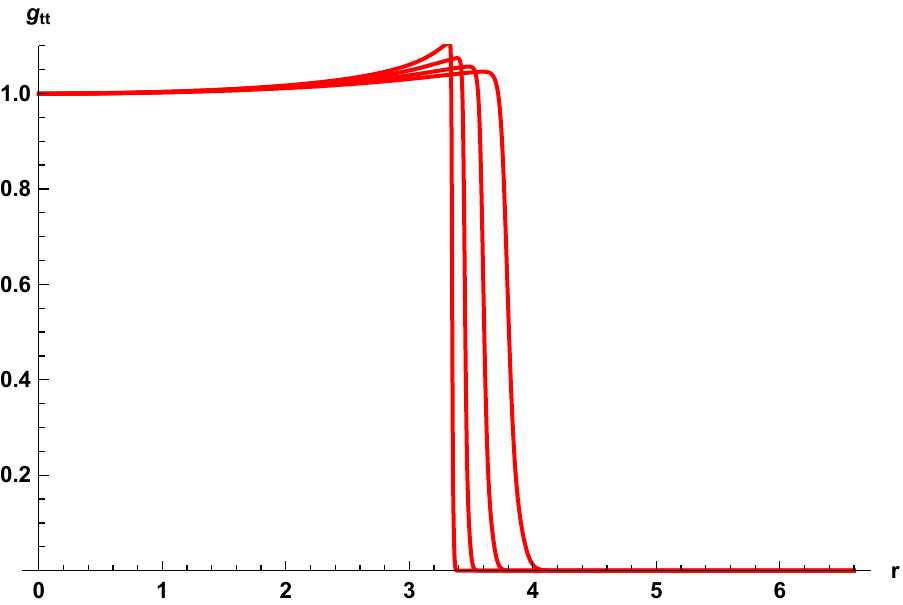} 
\caption{
Time component $g_{tt}(r,t)$ for $\rref=0$.
Left to right in $3.3<r<4$: $t=3$, 5, 7, 9.
}
\label{figgtt}
\end{figure}

\begin{figure}
\includegraphics[width=8cm]{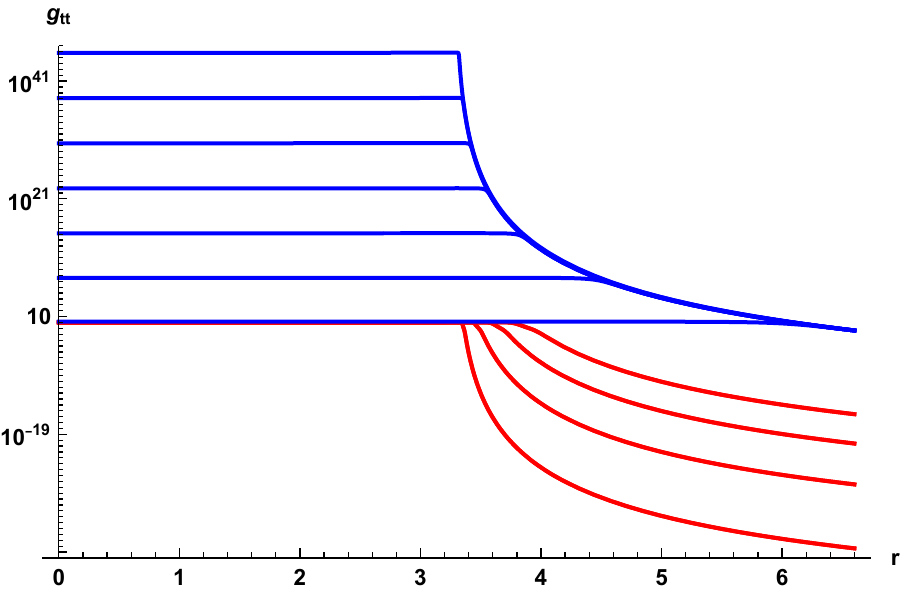} 
\caption{
Log-plot of $g_{tt}(r,t)$. The lower (red) curves correspond again to $\rref=0$, the upper (blue) curves to $\rref=2h$.
}
\label{figgttlog}
\end{figure}

The time component of the metric, $g_{tt}$ calculated from (\ref{gtata}) differs very much from $1/g_{rr}$ as shown in figures \ref{figgtt} and \ref{figgttlog}, note the vertical scale in the latter. The non-invariance of $g_{tt}$ under transformations of the time variable,  $t\to \bar t(t)$, has drastic effects here.

\begin{figure}
\includegraphics[width=8cm]{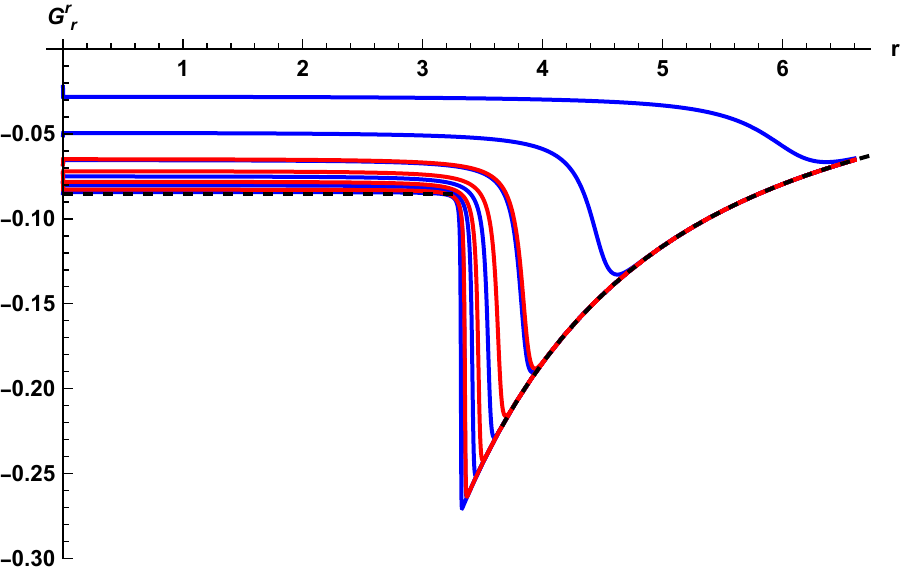} 
\caption{
Component $\Grr$. The black-dashed lines represent (\ref{Grr}), (\ref{Grrin}).
}
\label{figpGrr}
\end{figure}

\begin{figure}
\includegraphics[width=8cm]{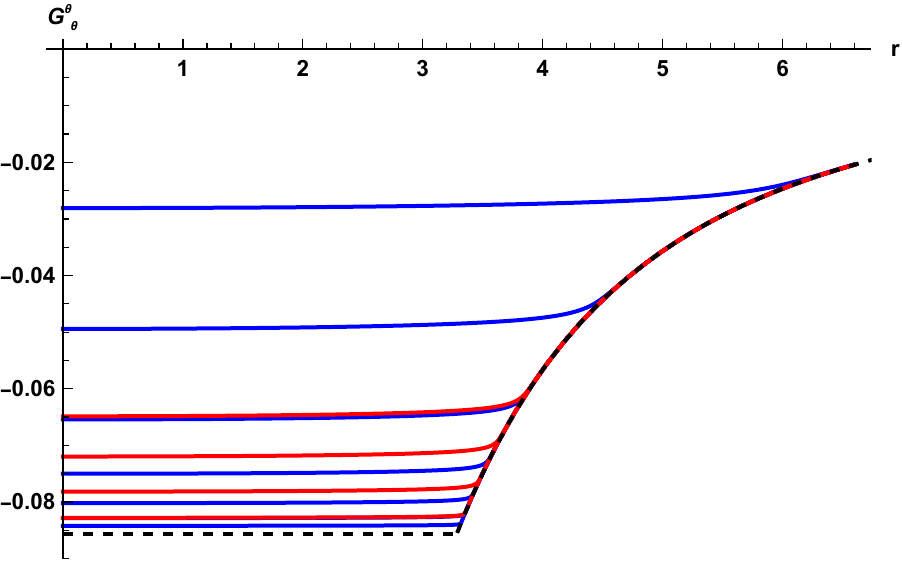} 
\caption{
Component $\Gthth$.  The black-dashed lines represent (\ref{Gtt}), (\ref{Grrin}).
}
\label{figpGthth}
\end{figure}

For the calculation of the components of the Einstein tensor the expressions given in Exercise 14.16 of \cite{Misner:1973prb} can be used; transformed to the Euclidean case they are recorded in appendix \ref{appET}.
Figure \ref{figpGtt} shows $\Gtata$ with the same conventions as in figures \ref{figfol} and following. It illustrates that $\rref=0$ and $\rref=2h$ give indeed the same $\Gtata$ curves as expected from the scalar nature of $\Gtata(r,t)$ under transformations $t\to \bar t(t)$. An envelope develops as $t\to 0$ which has a discontinuous jump at $r=h$. Figure \ref{figpGrr} shows similar phenomena in $\Grr$; for $\Gthth$ in figure \ref{figpGthth} the envelope is continuous ($\Gphph=\Gthth$). The Einstein tensor has an off-diagonal component $\Gtar$ shown in figure \ref{figpGtr}, which vanishes in the limit $t\to 0$ for $r\ne h$ and, as argued in appendix \ref{appEH}, also at $r=h$ when interpreted in the distributional sense.

\begin{figure}
\includegraphics[width=8cm]{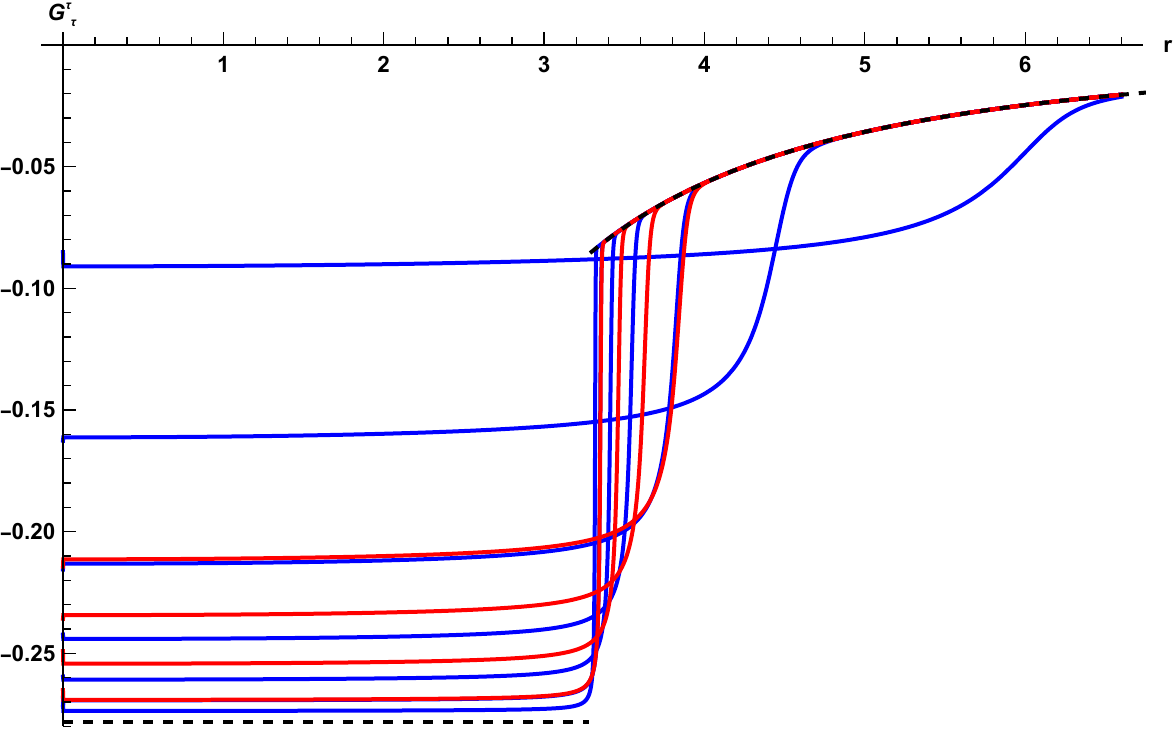} 
\caption{
Component $\Gtata$. Upper curves (blue) in $r<h$ correspond to $\rref=2h$, the curves occasionally joining (red) correspond to $\rref=0$. The black-dashed lines represent (\ref{Gtt}), (\ref{Gttin}), the limit $t\to 0$.
}
\label{figpGtt}
\end{figure}

\begin{figure}[t]
\includegraphics[width=8cm]{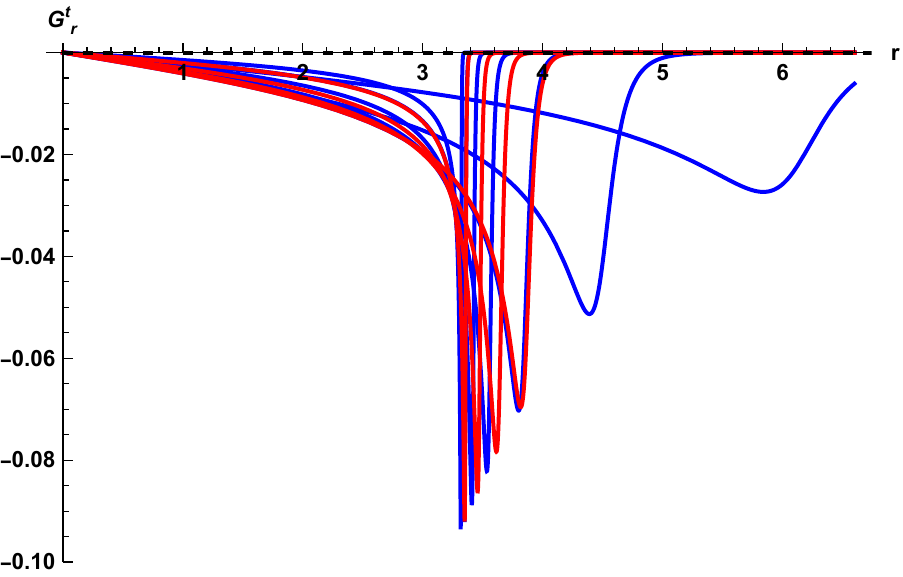} 
\caption{
Component $\Gtar$. It vanishes for $r\neq h$ in the limit $t\to 0$.
}
\label{figpGtr}
\end{figure}

The scalar curvature $R$ provides a check on the calculations. The trace of the Einstein tensor is $-R$. Calculating $R=6(-a^{\prime\prime}/a - a^{\prime 2}/a^2 + 1/a^2)$ from the metric (\ref{4D}), which is a function of $\et^2$,
using $\et^2(a)$ in (\ref{eta}) with $a=\sqrt{r^2+ \gf^2}$ and the numerically calculated $\gf(r,t)$ to transform it to $(r,t)$ variables, we can compare it with $-\Gmumu(r,t)$. The two $R$ match accurately with a point-wise precision of order $10^{-14}$ \% (using Mathematica with default working conditions). This is surprising, since one expects from (\ref{dotsy}) that the time derivatives of $\gf$ will have only order percent accuracy. The reason is that $\dot\gf$ and $\ddot\gf$ cancel out of the Einstein tensor, cf.\ appendix \ref{appET}. The shape of $\Gmumu$ similar to that of $\Gthth$.

\subsection{Singular transformation implementing $g_{tt}=g^{rr}$ at $t=0$}
\label{secttozero}

The numerical integration of $\gf'= P$ in section \ref{secnumerical} yielded diagonal metrics with spatial inverse components $1/g_{rr}=g^{rr}$ that approached a robust envelope when $t\to 0$.
On the contrary, the magnitude of the component $g_{tt}$ was highly sensitive to differing choices of boundary conditions (the two pole choices)
and the shape of $g_{tt}(r,t)$ differed from $g^{rr}(r,t)$.

In this section we seek to obtain $g_{tt}=g^{rr}$ in the limit $t \to 0$. We give up diagonality of the metric for general times but seek to recover it at zero time..
Consider erecting a pole as in the previous section but here at every $r\in(0,\rmax)$.  We wish to solve $\dot\gf = \sqrt{F}$ analytically at small times and try $\gf(r,t) = \fref(t;r)$.
It is helpful to take a brief look at the solution of this equation in case of the cosh-model,
\be
a(\et) = h \cosh(\et/r_0)\, ,
\label{coshmodel2}
\ee
which is simpler to solve than the rat-model.
(This model fitted the simulation data less accurately and failed at smaller Newton couplings \cite{Smit:2023kln}.
It approaches the exponential form of hyperbolic space at large $\et$, but here $r_0$ is meant to parametrize primarily the small to intermediate distance region. In particular, $1/r_0^2$ may be thought to represent $2(p-q)$ of the rat-model to $\mO(\et^2)$.)
With $a^\prime(\et)^2= (h^2/r_0^2) (\cosh^2(\et/r_0^2) -1)$ the function $F$ in (\ref{introF}) turns out as
\be
F(\sqrt{r^2 + \gf^2})=( r^2 + \gf^2 -h^2)/r_0^2 \,.
\label{Fcosh}
\ee
The solution of (\ref{fdot}$^+$) with initial condition (\ref{initialcond})$|_{\rref\to r}$ is
\bea
\gf(r,t) &=& \pm \sqrt{h^2-r^2}\,\cosh(t/r_0)\,,\; 0< r < h\,,\;
t \stackrel{\textstyle >}{<} 0\,,
\nonumber 
\\&=& \sqrt{r^2-h^2}\,\sinh(t/r_0)\,,\quad\;\;\; r> h\,.
\label{fsln}
\eea
At $t=0$ this solution satisfies $\gf'=P$ (trivially in the exterior as $0=0$), and the resulting metric becomes indeed diagonal, $g_{rt}=0$, with
\bea
g_{tt}(r,0) = 1/g_{rr}(r,0) &=& 1-r^2/h^2\,,\quad r<h\,,
\label{gt0}
\\
&=& (r^2 - h^2)/r_0^2= F(r)\,,\quad r>h\,.
\nonumber
\eea

Generalizing, in the exterior, consider a factorized form as suggested by  (\ref{fsln}),
\be
\gf(r,t) = u(r)\, t\,.
\label{yut}
\ee
Equation
(\ref{grta}) shows that it leads to a vanishing off-diagonal component at $t=0$, $g_{rt}(r,0)=0$, independent of $u(r)$.
(The equation $\gf'= P$ is again solved trivially as $0=0$.)
Furthermore (\ref{grr}) gives
\be
g_{rr}(r,t)= \frac{1}{F(r)} + \mO(t^2)\,,
\label{grrF}
\ee
also independent of $u(r)$, whereas (\ref{gtata}) gives
\be
g_{tt}(r,t) = u(r)^2 + \mO(t^2)\,.
\label{gtatauout}
\ee
Hence, requiring $g_{tt}(r,0)\, g_{rr}(r,0)=1$ we get
\be
u(r)=\sqrt{F(r)}\,.
\label{uout}
\ee
and the equation $\dot \gf = \sqrt{F}$ is indeed satisfied at $t=0$.

In the interior the Ansatz is,
\be
\gf(r,t) =\pm \sqrt{h^2 - r^2}\,(1+c_t\, t^2)\,,\quad
t \stackrel{\textstyle >}{<} 0\,,
\label{yutin}
\ee
which has a time-dependence similar to that of the cosh-model for small $t$
(cf.\ first line in (\ref{fsln})). Since $F[h]=a'(0)^2=0$ (cf.\ (\ref{introF})), $F$ vanishes in the interior at $t=0$:
\be
F(\sqrt{r^2 +\gf(r,0)^2})=F[h]=0\,.
\ee
Its expansion in $t$ starts out as
\be
F(\sqrt{r^2 +\gf(r,t)^2})=(h-r^2/h)\,F'(h)\, c_t\, t^2 + \mO(t^4)\,.
\ee
The equation $\gf'= P$ is satisfied at $t=0$. When $t\to 0$, the off-diagonal part of the metric $g_{rt}$ in (\ref{grta}) contains $1/F$ which blows up, and it contains $\dot\gf$ which vanishes. Working out the details we find
\be
g_{rt}= - 2 r \left(1+\frac{2}{h F'(h)}\right)\,c_t\, t + \mO(t^3)\,.
\ee
Hence, the interior metric becomes also diagonal at time zero. Furthermore, after some algebra:
\bea
g_{rr}(r,t) &=& \frac{h^2}{h^2-r^2} +\mO(t^2)\,,
\label{grrin}
\\
g_{tt}(r,t) &=& \frac{h^2-r^2}{h^2}\, \frac{2 h\, c_t}{F'(h)} + \mO(t^2)\,.
\label{gtatactin}
\eea
Requiring $g_{tt}(r,0)\,g_{rr}(r,0) =1$ determines $c_t$,
\be
c_t=\frac{F'(h)}{2 h}\,,
\label{ctF}
\ee
and with this choice the equation $\dot\gf=\sqrt{F}$ is indeed solved to leading order in $t$.
For the rat-model this becomes
\be
c_t = p-q\,,
\label{ctpq}
\ee
which was used in section \ref{secnumerical} for the case $\rref=0$.

Examples of $g_{tt}(r,0)$ following from (\ref{gtatauout}), (\ref{uout}),
(\ref{grrin}) -- (\ref{ctF}), have already been shown in figure \ref{figgttS1240}.

The Einstein tensor $\Gmunu$ contains time derivatives of the metric which is non-diagonal at non-zero $t$ and a little complicated. For the daunting task of its calculation the software OGRe \cite{Shoshany:2021iuc} is very helpful. The results in the limit $t\to 0$ are fairly simple and already given in \cite{Smit:2023kln}. For completeness we list the components again in the simplified notation (\ref{apq});   $\Gmunu(r,0)$ is diagonal and
in the exterior region, $r>h$:
\begin{widetext}
\bea
\Gtata = G^\theta_{\;\,\theta}=G^\ph_{\;\,\ph} &=&
-\frac{1}{r^2}\,\left[1 +
\frac{4 (h p - q r)^2 (h^2 p - 2 h(p + 2 q )\,r + 5 q\, r^2)}{h^2 (p - q )^2}\right]\,,
\label{Gtt}\\
G^r_{\;\, r} &=& -\frac{3}{r^2} \left[1 +
\frac{4 (h - r)(h p - q r)^3}{h^2 (p - q )^2})\right]\,.
\label{Grr}
\eea
\end{widetext}
In the interior region $0<r<h$:
\bea
\Gtata&=& -\frac{3}{h^2}\,,
\label{Gttin}
\\
G^r_{\;\, r} = G^{\ph}_{\;\,\ph}=G^{\theta}_{\;\,\theta}
&=& -\frac{1}{h^2}+4 (p - q) \,.
\label{Grrin}
\eea
These limit forms are shown in figures \ref{figpGtt} -- \ref{figpGtr} as the black-dashed curves. Note that $\Gtata$ and $\Grr$ are discontinuous at $r=h$, but not $\Gthth$.

In \cite{Smit:2023kln} it was conjectured that the discontinuity at $r=h$ in the first derivative of $g_{tt}(r,0)$ would be accompanied by a delta-shell $\propto \dl(r-h)$ in the transverse pressure $\propto\Gthth$, since the latter contains two derivatives of the metric with respect to $r$. If so, then one expects some regulated version of the Dirac distribution blowing up at small diminishing $t$. However, there is no sign of this in figure \ref{figpGthth}. Indeed, as shown in appendices \ref{appET} and  \ref{appEH}, such developing singular behavior is canceled exactly by a similar contribution with two time derivatives of the metric. After all cancelations are taken into account the Einstein tensor can be expressed in a form that depends explicitly only on $r$, $\gf$, $F$ and its first derivative $F'$, but not anymore on derivatives with respect to $r$ and $t$, cf.\ (\ref{Gttapp}) -- (\ref{Einstein}). These formulas apply equally well to the previous section as here and it seems remarkable that envelope in figure \ref{figpGthth} is described correctly by the consequences of (\ref{yut}) and (\ref{yutin}).

\section{Transforming the Hayward model}
\label{secH}

\begin{figure}
\includegraphics[width=8cm]{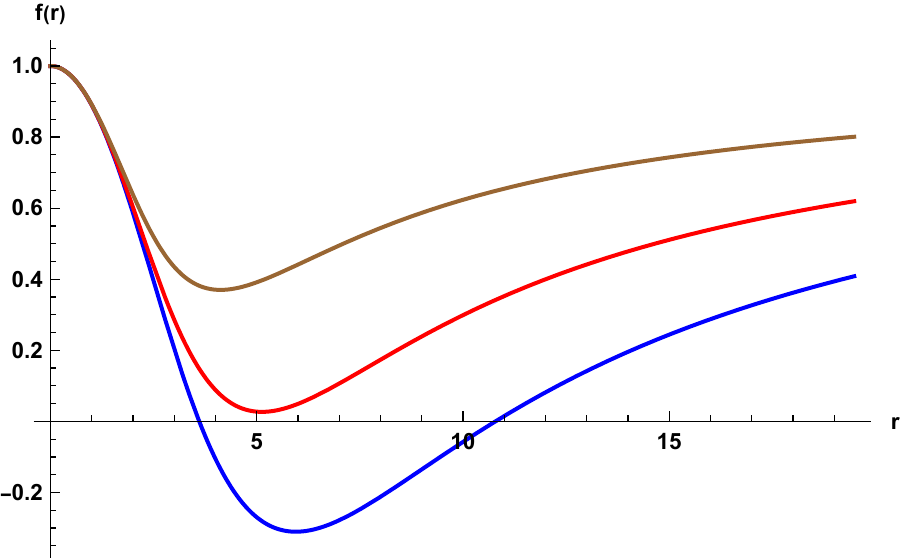} 
\caption{
Hayward function $f(r)$: top to bottom $m=0.5\,m_*$, $0.96\, m_*$, $1.5\, m_*$ and $l=3$.
}
\label{figfH}
\end{figure}

The  time-independence of the static version of Hayward's model \cite{Hayward:2005gi} allows a simple transformation to imaginary time, resulting in the Euclidean model:
\bea
ds^2 &=& f(r)\,dt^2 + \frac{1}{f(r)}\, dr^2 + r^2 \,d\Om_2^2\,,
\nonumber\\
f(r) &=& 1-\frac{2 m r^2}{2m l^2 + r^3}\,.
\label{defH}
\eea
The two parameters $m$ and $l$ are positive and have the dimension of length. For small $r$ the metric is like EdS, $f(r)\simeq 1-r^2/l^2$, and for large $r$ it is `Newton-like' $f(r)\simeq 1-2m/r$. Figure \ref{figfH} shows $f(r)$ for several values of $m$ at fixed $l$. With sufficiently large $m$ there are two real zeros at $r_+ > r_- >0$, which coincide, $r_+=r_-=r_*$, when $m$ is reduced to a critical value $m_*$. Then also $f'(r_*)=0$, which together with $f(r_*)=0$ determines
\be
m_* = \frac{3\sqrt{3}}{4}\,l\,,\quad r_*=\sqrt{3}\,l\,.
\ee

The aim here is to determine the scale factor $a(\et)$ in an $SO(4)$ invariant metric of the form (\ref{4D}) such that a transformation to spherical co\"{o}rdinates (\ref{spherical}) gives the metric (\ref{defH}) in the limit $t\to 0$.  For this purpose the results of the previous sections which led to $g^{rr}(r,0)=F(r)$ can be used (cf.\ (\ref{yut}) -- (\ref{uout})). Thus
\be
F(r) = f(r)\,,
\ee
and we can find $a(r)$ from the definition of $F$ in (\ref{introF}), by solving the differential equation
\be
a'(\et) =
\sqrt{f(a(\et))}\,.
\label{deqaet}
\ee
Since the Hayward model describes a regular black hole it is natural to fix the integration constant by requiring regularity at the origin,
\be
a(0)=0\,.
\label{bcaet}
\ee
To avoid a conical singularity also unit slope $a'(0)=1$ is required, which is satisfied since $f(0)=1$.

The small and large $\et$ behavior of the solution of (\ref{deqaet}), (\ref{bcaet}) are given by
\bea
a(\et) &\approx& l\,\sin(\et/l), \qquad\qquad\qquad\;\;\; a(\et)^3 \ll 2ml^2\,,
\\
&\approx& \et -m\ln(\et/m) +\mbox{const}, \quad a(\et)^3 \gg 2ml^2\,.
\label{aas}
\eea

When $m\geq m_*$,  the solution of (\ref{deqaet}), (\ref{bcaet}) has a fixed point at the first zero of $f(a)$,
\be
a(\et)\to r_-\,, \quad \et\to\infty, \quad (m\geq m_*)\,.
\ee
For $m<m_*$ but close to $m_*$ the solution can have a flat region
$a(\et)\approx  r_*$ before the asymptotic behavior (\ref{aas}) sets in.
Hence $r_*$ is similar to $h$ in the rat-model. However, the surface gravities at $r_*$ vanish,
$\kp_\pm = \lim_{r\to r_*^\pm} f'(r)/2=0$.

\begin{figure}
\includegraphics[width=8cm]{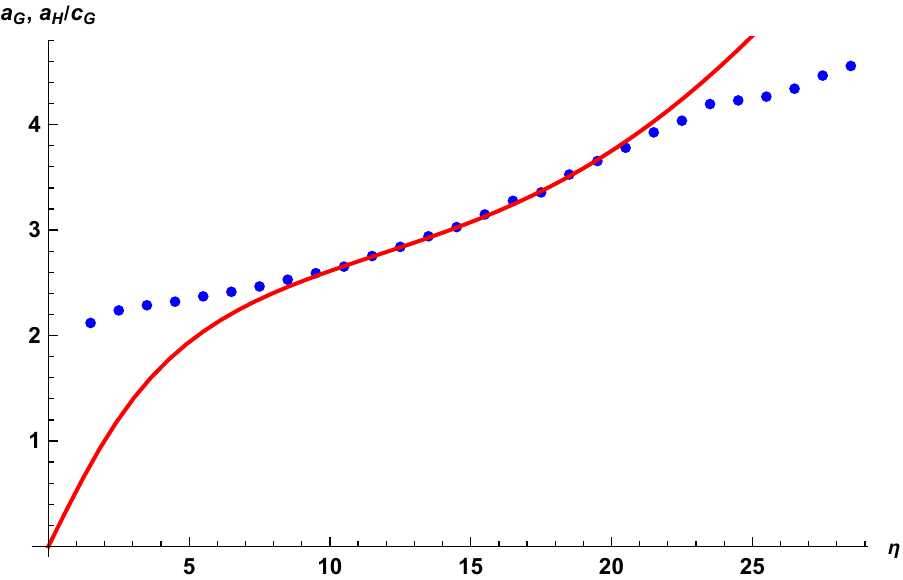} 
\caption{
Data of $a_G(\et)$ (blue) as in figure \ref{figFitaG} matched by $(1/c_G)\, a(\et)$, with $1/c_G=0.54$ and $a(\et)$ the solution of (\ref{deqaet}), (\ref{bcaet}) with $m=0.96\, m_*$, $l=3$.
}
\label{figFitaGH}
\end{figure}

A match (by hand and eye) of $(1/c_G)\, a(\et)$ to the data in figure \ref{figFitaG} is shown Figure \ref{figFitaGH}. Starting from the origin the curve reaches the data at much larger distances (by factor about 3) than in figure \ref{figFitaG}. This suggests that such fits may be more appropriate in computations reaching smaller lattice spacings ({\em in  physical units}).

\section{Conclusions}
\label{secconc}

We studied the effect of two types of co\"{or}dinate transformations,
a regular one producing a diagonal metric with $g_{tt}(r,t)\neq g^{rr}(r,t)$ at all times and a singular one \cite{Smit:2023kln} producing a metric which becomes diagonal only in the limit $t\to 0$ with $g_{tt}(r,0)=g^{rr}(r,0)$.  At time zero the component $g^{rr}(r,0)$ is the same for both transformations As shown in figure \ref{figgttS1240} it is zero at $r=h$, as for a black hole with gravitational radius $h$. The singular transformation has a singularity at this point, which is akin to the transformation between the Schwarzschild-Droste black hole and its Kruskal-Szekeres extension
\cite{Schwarzschild:1916uq,Droste:2002,McGruder:2018bsi,Kruskal:1959vx,Szekeres:1960gm}.

The numerical study of the regular transformation showed that the left cuspy curve in figure \ref{figgttS1240} is the limiting envelope of smooth curves representing $g^{rr}(r,t)$ (cf.\ figure \ref{figginv}). The time component $g_{tt}(r,t)$ differed from $g^{rr}(r,t)$,  in shape and, depending on boundary conditions specifying the implementation of the transformation, also by many orders of magnitude.
These changes in boundary conditions correspond to transformations $t\to\bar t(t)$
under which $\gf(r,t)$ and $g^{rr}(r,t)$ behave as scalar fields, but not $g_{tt}(r,t)$ which may suffer large multiplicative factors. Diagonal mixed components of the Einstein tensor, $\Gmunu(r,t)$, $\mu=\nu$,  do also transform as a scalar field under $t\to\bar t(t)$ and they approach the same zero time limit as with the singular transformation. Remarkably, the off-diagonal $\Gtar(r,t)$ also does not suffer from the time ambiguity; it was found to vanish when $t\to 0$ and arguments where given that this holds also in the distributional sense.

In \cite{Smit:2023kln} a shell-like singular distribution was conjectured in the transverse pressure $\propto \Gthth(r,0)$ at $r=h$  as a consequence of the discontinuous first derivative $2\kp_\pm = \lim_{r\to h^\pm}\partial_r g_{tt}(r,0)$.
However, with the regular transformation such a shell is absent due to a remarkable cancellation between singular contributions to $\Gthth$ (cf.\ appendix \ref{appEH}). One also notes that the local minimum in $g^{rr}(r,t)$ approaches $r=h$ when $t\to 0$ (figure \ref{figginv}), which
implies that the derivative at the minimum remains zero in the limit.
A gravastar-like shell \cite{Mazur:2015kia} is not supported in this study.

The inversely transformed Hayward model and its qualitative match to the EDT data sheds new light on the interpretation of the EDT results.

\appendix
\section{Einstein tensor for a time-dependent diagonal metric}
\label{appET}
In equation (14.48) of \cite{Misner:1973prb}, a time-dependent rotation-invariant diagonal Lorentzian metric depending on  co{\"o}rdinates $T$, $R$, $\theta$ and $\ph$ is defined by its line element
\be
ds^2 = -e^{2\Ph}\,dT^2 + e^{2\Lm}\, dR^2 + r^2 (d\theta^2 + \sin(\theta)^2 d\ph^2)\,,
\ee
and its Einstein tensor is given in (14.48), (14.51) and (14.52) of this book. Transforming co{\"o}rdinates $T=-i\,t$ and choosing $R=r$ we obtain the Einstein tensor for the Euclidean metric.
The transformed functions of \cite{Misner:1973prb} are given by
\be
\Ph = \half\,\ln g_{tt}\,,\quad \Lm = \half\,\ln g_{rr}\,,
\ee
and
\bea
E_1 &=& e^{-2\Ph} (\ddot \Lm + \dot\Lm^2 -\dot\Lm \dot \Ph)\,,
\\
E_2&=&-e^{-2\Lm} (\Ph''+\Ph'^2-\Ph^{\prime}\Lm')\,,
\\
E &=& E_1 + E_2\,,\quad \bar E = -\frac{1}{r}\,e^{-2\Lm}\; \Ph'\,,
\\
F_{\mbox{\tiny{MTW}}}&=&\frac{1}{r^2}\,(1-e^{-2\Lm})\,,\quad \bar F_{\mbox{\tiny{MTW}}} = \frac{1}{r}\,e^{-2\Lm}\; \Lm'\,,
\\
H&=& -i \frac{1}{r}\,e^{-\Ph-\Lm}\; \dot\Lm\,,
\eea
in terms of which the Euclidean components of the Einstein tensor are given by
\bea
G^t_{\;\,t} &=& - F_{\mbox{\tiny{MTW}}} - 2\bar F_{\mbox{\tiny{MTW}}}\,,\quad \Grr= -2\bar E - F_{\mbox{\tiny{MTW}}}\,,
\nonumber
\\
\Gthth &=&\Gphph = -E - \bar E - \bar F_{\mbox{\tiny{MTW}}} \,,\quad \Gtar= 2\, i H\,.
\label{EGMTW}
\eea
Note that $E_1$ contains two time derivatives of $g_{rr}$, $E_2$ two spatial derivatives of $g_{tt}$, and that their sum $E_1+E_2$ enters in $\Gthth$.
It is convenient to treat $F(a)$ introduced in (\ref{introF}) as a function of $a^2=r^2 + \gf^2$, writing
\be
F(\sqrt{r^2 + \gf^2}) = \Ft(r^2 + \gf^2)\,.
\label{introtF}
\ee
The functions $E$, \ldots, $H$ depend on derivatives of the metric components
\be
\fg \equiv g_{tt}\,,\quad \fgi \equiv \frac{1}{g_{rr}}
\ee
(cf.\ (\ref{gtata}), (\ref{grr})). Indicating their explicit dependence on variables these are given by
\bea
\fgi(r,\gf)&=& \frac{r^2 \Ft(r^2 + \gf^2) + \gf^2}{r^2 + \gf^2}\,,
\\
\fg(r,\gf,\dot\gf) &=& \frac{\dot\gf^2}{\Ft(r^2 + \gf^2)}\,\fgi(r,\gf)\,.
\eea
Since $\gf$ is a solution of $\partial_r\gf=P$,
\be
P(r,\gf)= r\gf \frac{-1+\Ft(r^2 + \gf^2)}{\gf^2 + r^2 \Ft(r^2 + \gf^2)}\,,
\ee
we can replace spatial derivatives $\partial_r\gf$ by $P$:
\bea
\frac{d}{dr}\,\fgi &=& \frac{\partial \fgi}{\partial r} + \frac{\partial \fgi}{\partial \gf}\, P\,,
\\
\frac{d}{dr}\,
\fg &=& \frac{\partial \fg}{\partial r} + \frac{\partial \fg}{\partial \gf}\, P
+ \frac{\partial \fg}{\partial \dot\gf}\, \partial_r \dot\gf\,,
\\
\partial_r\,\dot\gf
&=&\partial_r\partial_t \gf
= \partial_t \partial_r \gf =
\frac{d}{d t}\, P = \frac{\partial P}{\partial \gf}\, \dot \gf\,,
\eea
and recursively for higher derivatives. In principle this gives $E_1$, $E_2$, \ldots, $H$ as functions depending explicitly only on $r$, $\gf$, $\dot\gf$, $\ddot\gf$ ($\dddot\gf$ is not needed since double time derivatives occur only on $\Lm$ which does not contain $\dot\gf$). However the explicit dependence on  $\dot\gf$ and $\ddot \gf$ cancels out. This should happen in the diagonal components of the Einstein tensor which transform as a scalar field under time transformations $\bar t(t)$ and $\dot\gf$ and $\ddot\gf$ are not such scalars. But it happens already in the individual $E_1$, $E_2$, \ldots, $H$.

Further details of $E_1$, \ldots, $H$ are in appendix \ref{appEH}.
There is a near cancellation between $E_1$ and $E_2$, hence also in their sum $E$ contributing to $\Gthth$. This is relevant because $E_1$ and $E_2$ separately are increasingly strongly peaked as $t\to 0$, whereas $E$ clearly reaches a finite limiting form.
The  Dirac distribution $\dl(r-h)$, which was conjectured in \cite{Smit:2023kln} to be present in $\Gthth$ at $t=0$ as a result of the double spatial derivative in $E_2$, is not present in $E$.   We were not able to prove that $\dl(r-h)$ emerges in $E_1$ and $E_2$ separately as $t\to 0$, although there is modest numerical evidence for a finite limit in the distributional sense.

The components of $\Gmunu$ listed in (\ref{EGMTW}) become:
\bea
\Gtata&=& \frac{(\Ft-1)(3 \gf^2 + r^2 \Ft)}{(\gf^2 + r^2) (\gf^2 + r^2 \Ft)}
+ \frac{2 r^2 \Ft \Ft'}{\gf^2 + r^2 \Ft}\,,
\label{Gttapp}
\\
\Grr &=&\frac{(\Ft-1)(\gf^2 + 3 r^2 \Ft)}{(\gf^2 + r^2) (\gf^2 + r^2 \Ft)}
+\frac{2 \gf^2 \Ft'}{\gf^2 + r^2 \Ft}\,,
\label{Grrapp}
\\
\Gthth &=& \Gphph =\frac{\Ft-1}{\gf^2 + r^2} + 2 \Ft'\,,
\label{Gththapp}
\\
\Gtar &=& 2 r \gf \sqrt{\Ft}\left[\frac{\Ft-1}{(\gf^2 + r^2) (\gf^2 + r^2 \Ft) }
- \frac{\Ft'}{\gf^2 + r^2 \Ft}\right]\,.
\nonumber
\\
\label{Einstein}
\eea
Here $\Ft'$ is the derivative of $\Ft$ with respect to its argument: $\Ft'(r^2 + \gf^2)=d \Ft(x)/dx|x\to r^2 + \gf^2$ (and similar for $\Ft^{\prime\prime}$ which appears in $E_1$ and $E_2$). The trace of the Einstein tensor simplifies to
\be
\Gmumu = \frac{6 (-1 + \Ft)}{\gf^2 + r^2} +  6\, \Ft^\prime \,.
\ee
Following the reasoning in section \ref{secttozero}, the limit forms for $t\to 0$ in that section follow here easily -- without encountering singularities -- by letting $\gf\to 0$ in the exterior region and
$\Ft\to 0$ \& $\gf\to \sqrt{h^2 - r^2}$ in the interior region. The off-diagonal component $\Gtar$ vanishes in the limit. The possibility of a remaining finite distribution at $r=h$ is investigated in appendix \ref{appEH}.

\section{Details of $E_1$, \ldots, $H$}
\label{appEH}

After the canceling-out of $\dot\gf$ and $\ddot\gf$, $E_1$ and $E_2$ are given by
\bea
E_1 &=& e_{10} + e_{11}\,\Ft'+ e_{12}\, (\Ft^{\prime})^2\,,
\nonumber\\
E_2 &=& e_{20} + e_{21}\, \Ft' + e_{22}\, (\Ft^{\prime})^2\,,
\nonumber\\
e_{10} &=& \frac{1}{(y^2 + \Ft r^2)^3}\left[r^2 (3 y^2 (-1 + \Ft) \Ft - (-1 + \Ft) \Ft^2 r^2)\right.
\nonumber\\
&&\left. +
 \Ft^{\prime\prime} r^2 (2 y^6 \Ft + 2 y^4 \Ft (1 + \Ft) r^2 + 2 y^2 \Ft^2 r^4)\right]\,,
 \nonumber\\
 e_{11} &=& \frac{1}{(y^2 + \Ft r^2)^3}\left[r^2 (\gf^4 (1 - 4 \Ft) + \Ft^2 r^4 \right.
 \nonumber\\
 &&\left.  + \gf^2 \Ft (-6 r^2 + 4 \Ft r^2))\right]\,,
 \nonumber\\
 e_{12} &=& \frac{1}{(y^2 + \Ft r^2)^3}\left[r^2 (\gf^6 + \gf^4 (1 - 3 \Ft) r^2 - 3 \gf^2 \Ft r^4)\right]\,,
 \nonumber\\
 e_{21}&=&  \frac{1}{(y^2 + \Ft r^2)^3}\left[(-\gf^6 + \gf^4 (-1 + \Ft) r^2 \right.
 \nonumber\\
 &&\left. + \gf^2 \Ft r^2 (6 r^2 - 7 \Ft r^2) - \Ft^2 r^4 (r^2 + \Ft r^2))\right]\,,
 \nonumber\\
 e_{20} &=& -e_{10}\,,\quad e_{22}=-e_{12}\,,\quad
 e_{11} + e_{21} = -1\,,
 \nonumber\\
 E &=& E_1 + E_2 = -\Ft'\,.
 \eea
 The expressions for  $\bar E$, \ldots, $H$ come out as:
\bea
\bar E &=& -\frac{(-1 + \Ft) \Ft r^2}{(\gf^2 + r^2) (\gf^2 + \Ft r^2)}
-\frac{\gf^2 \Ft^\prime}{\gf^2 + \Ft r^2} \,,
\nonumber\\
F_{\mbox{\tiny{MTW}}} &=& \frac{1 - \Ft}{\gf^2 + r^2}\,,
\nonumber\\
\bar F_{\mbox{\tiny{MTW}}} &=& -\frac{\gf^2 (-1 + \Ft)}{(\gf^2 + r^2) (\gf^2 + \Ft r^2)}
-\frac{\Ft \Ft^\prime r^2}{\gf^2 + \Ft r^2}\,,
\nonumber\\
i H &=&r \gf\sqrt{\Ft}\left[\frac{\Ft-1}{(\gf^2+r^2)(\gf^2 + \Ft r^2)} -\frac{\Ft^\prime}{\gf^2 + \Ft r^2}\right]\,.
\eea

For the rat-model $\Ft$ and $\Ft'$ can be written in the form (cf.\ (\ref{Frat}))
\bea
\Ft&=&c (\sqrt{x}-h) (\sqrt{x} - \bar h)^3\,,\quad x=r^2 + \gf^2\,,
\\
\Ft'&=& \frac{c (4\sqrt{x}-3 h - \bar h)(\sqrt{x}-\bar h)^2}{2\sqrt{x}}\,.
 \eea

\begin{figure}
\includegraphics[width=8cm]{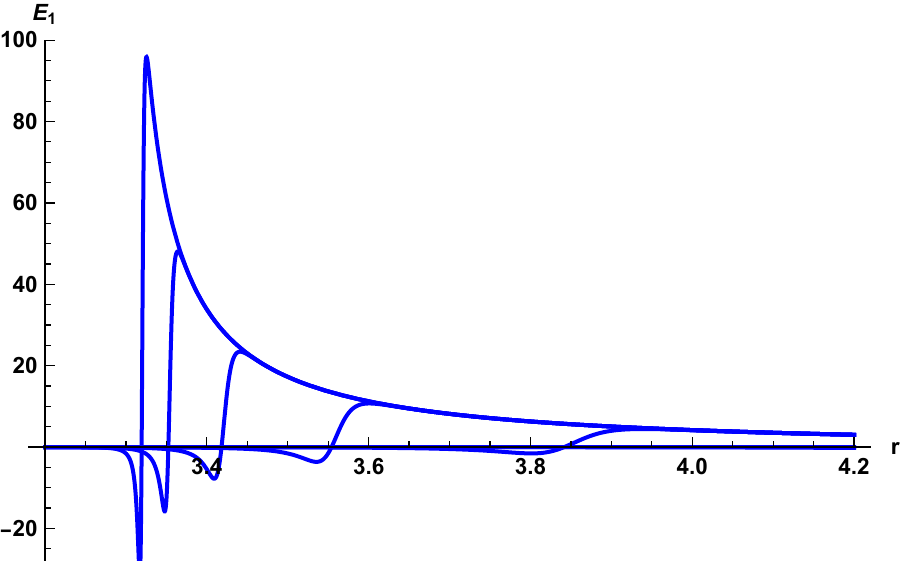} 
  \caption{Plot of $E_1(r,t)$ for $\rref=2 h$ and $t= 10^{k-24}$, $k=0,4,8,12,16$.
  }
\label{figE1}
\end{figure}

\begin{figure}
\includegraphics[width=8cm]{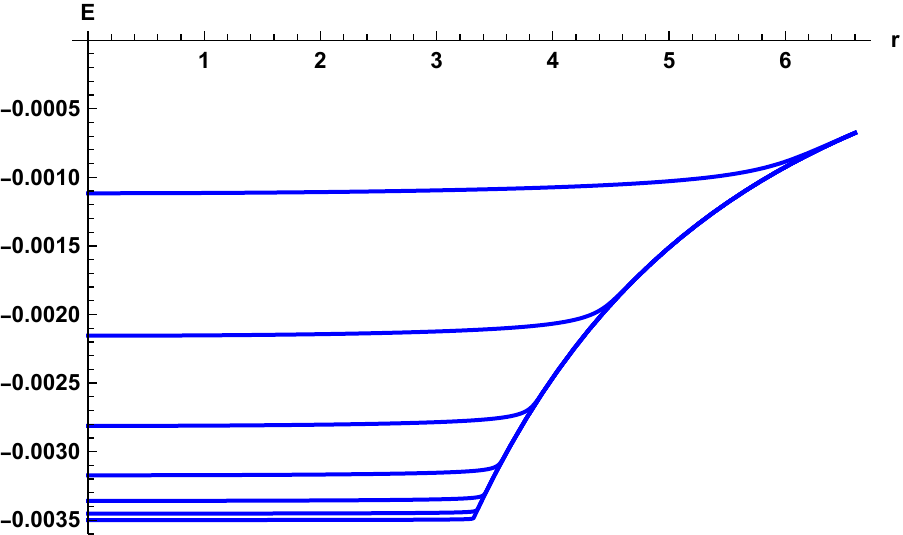} 
  \caption{Plot of $E=E_1+E_2$ for the same times as in figure \ref{figE1}.
  }
\label{figE}
\end{figure}

Figure \ref{figE1} shows a plot of $E_1$ corresponding to five (blue) curves of $\gf(r,t)$ in figure \ref{figfol} with $\rref=2h$.  A similar plot for $-E_2(r,t)$ is indistinguishable to the eye, since the sum $E_1 + E_2$ is down in magnitude by a factor of about $10^4$, note the vertical scale in figure \ref{figE} which displays $E$.

\begin{figure}
\includegraphics[width=8cm]{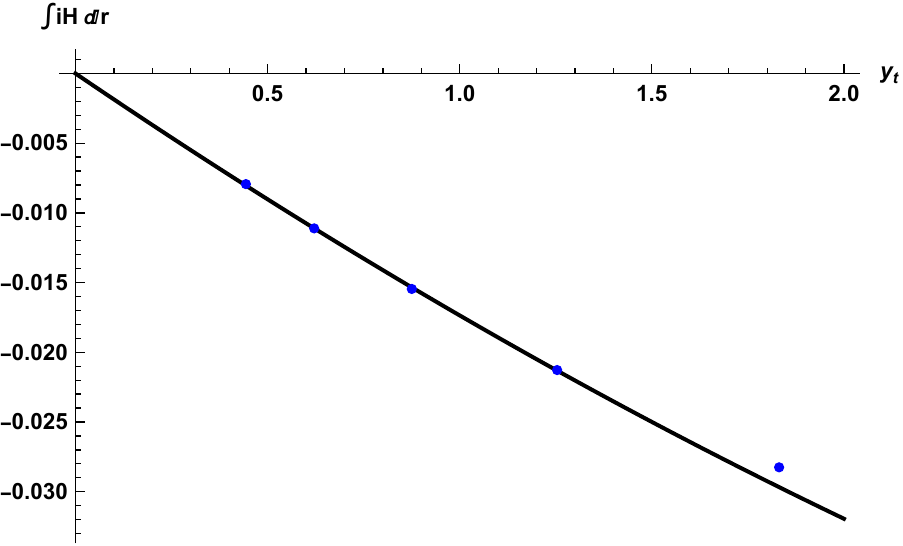} 
  \caption{Plot of $\int_0^{2h} dr\, iH$ vs.\ $\gf_t$ for the same times as in figure \ref{figE1}. The black line represents a fit by $-0.0187\, \gf_t + 0.00139\, \gf_t^2$ to the first four values.
  }
\label{figpintH}
\end{figure}

The integral $I=\int_{3.2}^{4.2} dr\, E_1(r,t)$ was monitored to check wether a finite distribution (such as a Dirac function $\dl(r-h)$) develops in $E_1$ (and also in $E_2$ as follows from the cancelation) in the limit $t\to 0$. Considered as function of $\ln(t)$, $I$ is very well fitted by the form $I=\al + \bt \ln(t)$, which suggests a logarithmic divergence in the limit. But a dependence of $I$ as a function of a time $t$ introduced at $\rref=2h$, `far away' from $r=h$ involves a  co{\"o}rdinate peculiarity of this $t$ (cf.\ figure \ref{figfollog}). Testing as a function of the foliation as labeled by the value of $\gf$ at $r=h$, i.e.\ $\gf_t=\gf(h,t)$, may be a better idea. The values of $I$ are well fitted by the rational-function form
$I=(\al +\bt\, \gf_t)/(1+\gm\, \gf_t)$, which has a build-in finite limit as $\gf_t\to 0$. We take this as a mild support for a finite limit distribution $E_{1,2}$ at $r=h$.  However, since only the regular sum $E$ enters in $\Gthth$ the finiteness of $E_{1,2}$ is not of physical interest.

The function $2 i H=\Gtar$ has been plotted in figure \ref{figpGtr}.
It vanishes when $t\to 0$ for $r\neq h$ (as also mentioned in appendix \ref{appET}). To investigate the possibility of a finite remaining distribution at $r=h$, consider $I_H=\int_0^{2h} dr\, i H$. It turns out to be a non-linear function of $\ln(t)$ but an almost linear one as a function of $\gf_t=\gf(h,t)$; its four smallest values can be fitted by the form $\bt\, \gf_t + \gm\, \gf_t^2$, as shown in figure \ref{figpintH}.
Adding a constant $\al$ to the fit function leads to a rather small value $\al=0.00075$.
We assume that $H$ vanishes also as a distribution when $t\to 0$.\\

\bibliography{lit}

\end{document}